# Electrochemical Ion Insertion: From Atoms to Devices

Aditya Sood[1,2,*], Andrey D. Poletayev[1,2,*], Daniel A. Cogswell[3,*], Peter M. Csernica[2,*], J. Tyler Mefford[1,2,*], Dimitrios Fraggedakis[3], Michael F. Toney[4,5,#], Aaron M. Lindenberg[1,2,6,#], Martin Z. Bazant[3,7,#], and William C. Chueh[1,2,#]

[1]Stanford Institute for Materials and Energy Sciences, SLAC National Accelerator Laboratory, Menlo Park, CA 94025, USA
[2]Department of Materials Science and Engineering, Stanford University, Stanford, CA 94305, USA
[3]Department of Chemical Engineering, Massachusetts Institute of Technology, Cambridge, MA 02139, USA
[4]Stanford Synchrotron Radiation Lightsource, SLAC National Accelerator Laboratory, Menlo Park, CA 94025, USA
[5]Department of Chemical and Biological Engineering, University of Colorado Boulder, Boulder, CO 80309, USA
[6]PULSE Institute, SLAC National Accelerator Laboratory, Menlo Park, CA 94025, USA
[7]Department of Mathematics, Massachusetts Institute of Technology, Cambridge, MA 02139, USA

[*]These authors contributed equally.

[#]Corresponding authors:
mftoney@slac.stanford.edu, aaronl@stanford.edu, bazant@mit.edu, wchueh@stanford.edu

**Electrochemical ion insertion involves coupled ion-electron transfer reactions, transport of guest species, and redox of the host. The hosts are typically anisotropic solids with two-dimensional conduction planes, but can also be materials with one-dimensional or isotropic transport pathways. These insertion compounds have traditionally been studied in the context of energy storage, but also find extensive applications in electrocatalysis, optoelectronics, and computing. Recent developments in operando, ultrafast, and high-resolution characterization methods, as well as accurate theoretical simulation methods, have led to a renaissance in the understanding of ion-insertion compounds. In this Review, we present a unified framework for understanding insertion compounds across time and length scales ranging from atomic to device levels. Using graphite, transition metal dichalcogenides, layered oxides, oxyhydroxides, and olivines as examples, we explore commonalities in these materials in terms of point defects, interfacial reactions, and phase transformations. We illustrate similarities in the operating principles of various ion-insertion devices ranging from batteries and electrocatalysts to electrochromics and thermal transistors, with the goal of unifying research across disciplinary boundaries.**

To many, electrochemical ion-insertion solids are synonymous with Li-ion battery electrodes[1]. Indeed, materials that accommodate the insertion and removal of Li into and out of their structure through coupled ion/electron transfer redox reactions have found their most prevalent application in powering mobile electronic devices; widespread adoption of electric vehicles and the use of batteries to decarbonize the electric grid are also becoming realities. John Goodenough, Stanley Whittingham, and Akira Yoshino were awarded the 2019 Nobel Prize in Chemistry for their seminal work on developing this ubiquitous technology[2]. The story of electrochemical ion insertion materials, however, extends well beyond batteries. Early work carried out during the *chimie douce*, or "soft chemistry," movement of the 1980's and 90's established electrochemical ion insertion as a synthetic strategy to change the chemical composition of a host material through the addition of guest species into the interstitial sites of the stable host crystal structure[3–6]. Today, ion insertion plays a central role in numerous applications beyond batteries, such as electrocatalysis, electrochromics, thermal switching, separations, desalination, and neuromorphic transistors (**Figure 1**). Although these devices operate across a broad range of length and time scales, they share the same basic working principle: an applied electrical potential inserts ions into (or removes ions from) a host material, altering its local chemical, electronic, and lattice structure and modulating its optical, electronic, and thermal properties.

A typical ion-insertion device consists of two mixed ion-electron-conducting electrodes separated by an ion-conducting electrolyte. Ion insertion is driven by a gradient in the electrochemical potential of the ions being inserted, and is accompanied by a second, charge-compensating carrier entering the same material[7]. An applied external voltage/current enables precise control over the number of ions being inserted into the host material. Although ion insertion is conceptually simple, its underlying microscopic details are complex and not completely understood. At least three main processes are involved: the desolvation of the ion in the electrolyte (most well studied for liquid electrolytes), the interfacial transport of the ion from the electrolyte into the host lattice, and the transfer of an electron from the host lattice i.e. redox. For bulk insertion to occur, both an electron and an unoccupied site at the surface of the host lattice must be available. Once the ion has entered the host lattice from the electrolyte, it maintains its charge, while the electron is transferred to the host lattice, where it may become localized[8]. Thus, migration of the ion occurs via the ambipolar hopping of an ion-electron pair.

The idea that an electric field can be used to control material properties is not new. In fact, the "field effect" lies at the heart of the semiconductor revolution[9]. It is the process by which an electric field modulates the concentration of free carriers in a material through the action of electrostatic forces. Key to the field effect is the concept of doping, wherein impurities are incorporated into a pure material (typically silicon) in minute quantities. These impurities can either act as donors or acceptors, introducing conduction-band electrons or valence-band holes in the continuum. Importantly, once the semiconductor device has been fabricated, the concentration and distribution of dopants (typically on the order of parts per million) are fixed, or "quenched". An applied electric field changes only the local concentration of mobile carriers, but does not directly influence the ionized dopants, which are immobile.

In contrast to the field effect, ion insertion involves a different approach to tuning material properties. First, the concentration of inserted ions is not fixed, but is dynamically variable. Furthermore, the composition can be actively tuned over a very wide range, from a fraction of a percent up to one or more guest atoms per unit cell. This leads to a second important consequence, that the effects of insertion can no longer be considered in terms of a dilute and continuum picture. Strong local modifications occur in the bonding configuration and electronic structure of the host. Third, interactions between different species, such as between the inserted ion and host atoms or between the inserted ions themselves, result in higher order atomic structures that determine various parameters such as the activation energy for ionic transport. An essential consequence of all the above features is that ion insertion is a nonlinear phenomenon with respect to ion concentration: the large concentration of guest species, and their mutual interactions, causes intrinsic properties such as ionic mobility and point defect formation energy to become composition dependent.

In this Review, we briefly summarize the basic materials science principles behind this class of materials and demonstrate how electrochemical ion insertion can be used to tune material properties towards a wide variety of applications from mature technologies such as batteries and electrochromics to emerging ones in electrocatalysis, thermal regulation, neuromorphic computing, and water purification. The unique aspect of ion insertion that enables tuning of physical (optical, electronic, thermal etc.) and chemical properties is the simultaneous modulation of the host material's atomic and electronic structures, through the insertion of both ions and compensating electrons. Note that we do not focus on the pseudo-capacitance of fast-charging porous electrodes, which sometimes include ion-insertion nanoparticles, since this phenomenon has been recently reviewed in a broader context[10]. While the fundamental concepts discussed below are largely focused on crystalline inorganic materials, some of these ideas may apply more broadly to other mixed ion-electron conductors. These include amorphous organic materials like polymers[11], and crystalline organic-inorganic materials like hybrid perovskites[12]. For a pedagogical introduction to some of the concepts discussed here, the readers are suggested to review relevant textbooks[7,9,13–16]. It is not our goal to provide a thorough technological review of the various ion insertion applications, but rather to demonstrate the fundamental principles behind their operation. In doing so, we hope to unite the efforts of these different fields and present a framework for understanding electrochemical ion insertion across time and length scales ranging from the atomic to device level.

# Use cases

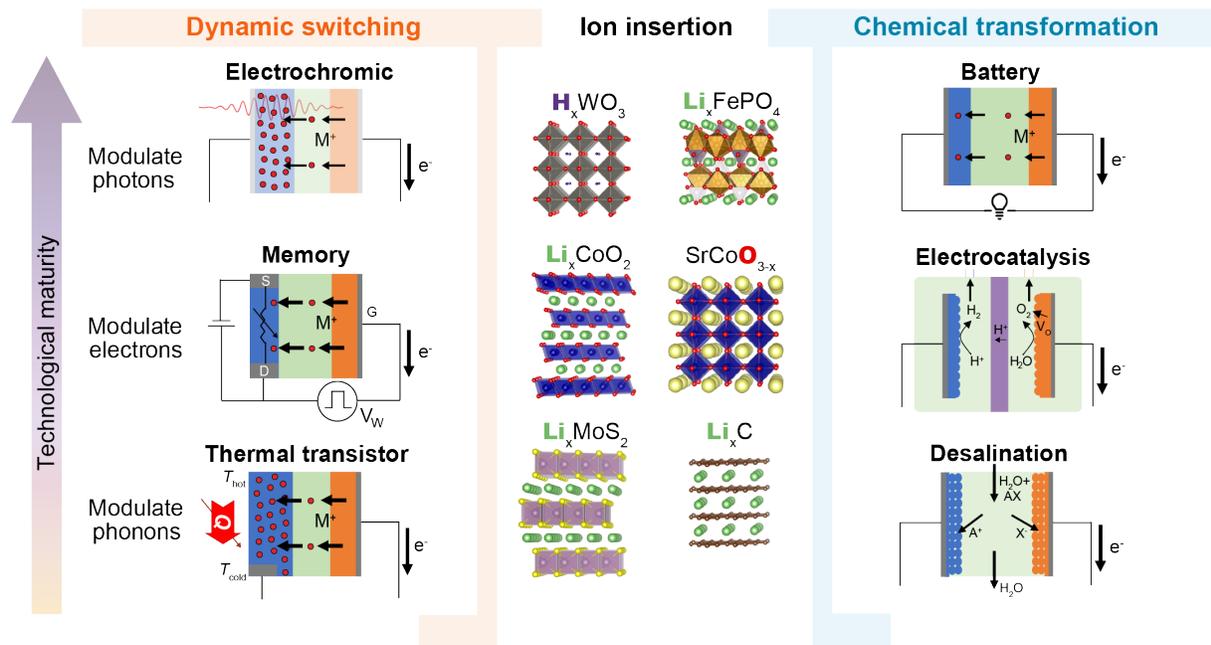

**Figure 1: Electrochemical ion insertion plays a key role in a wide variety of applications.** The first category harnesses ion insertion to create dynamic switching materials with actively tunable optical, electronic, and thermal properties, by modulating the transport of photons, electrons, and/or phonons through the host material (left panel). The second category utilizes the chemical transformations induced by ion insertion, for example, in energy storage, electrocatalysis, and desalination (right panel). In the device schematics, the blue and orange shaded regions are the electrodes, and green regions represent the electrolyte. Examples of commonly studied materials are shown in the center panel. These can, in general, be inorganic or organic. Often, the same material finds use in more than one type of ion insertion device.

Ion insertion devices are broadly classified into two categories, as shown in **Figure 1**. The first is where ion-insertion imparts functionality to a host material by dynamically switching its physical properties[17,18] (left panel of **Figure 1**). In the following section, we discuss how ion insertion enables active control over the three main modes of energy transport, namely light, charge, and heat, by tuning the host's interactions with photons, electrons, and phonons, respectively. In recent years, this line of research has led to the creation of "smart" windows that could improve building energy efficiency, brain-inspired artificial neurons, and thermal circuits that could enable heat-based computers. The second category includes applications where ion-insertion is used to control chemical transformations, as in the case of energy storage, electrocatalysis, and desalination (right panel of **Figure 1**). We will elucidate how ion insertion influences surface catalytic activity for technologically-relevant chemical reactions like water splitting, and discuss how it enables water purification and lithium harvesting through the selective extraction of ions. While batteries are arguably the most ubiquitous ion insertion devices, we do not dedicate a separate section to them. Instead, we will use examples from battery research throughout this Review to illustrate unifying fundamental principles.

## *Ion insertion for dynamic switching*

**Electrochromics: Modulate photons**

Among the ion-insertion based switching devices discussed in this review, electrochromics (ECs, materials with optical properties tunable by voltage) are by far the most advanced on the technological readiness scale. Applications of ECs include tinted windows in aircraft, antiglare mirrors in automobiles, and so-called smart windows in energy-efficient buildings. There are both economic and environmental drivers for this technology. For example, it is estimated that dynamic windows can reduce the amount of energy spent on lighting, heating, ventilation, and air-conditioning (HVAC) in buildings[19] by 10-20%, which account for nearly one-fifth of all U.S. energy consumption[20]. For a general overview of ECs, we refer the reader to previous reviews[21–25]; here, we focus on the working principles of ion-insertion based ECs.

A typical EC device consists of an electrochromic working layer whose optical properties change upon charge injection (and accompanying ion insertion), an ion storage layer that acts as a counter electrode, and an ion-conducting electrolyte. The EC and ion storage layers are mixed ion-electron conductors. For smart window applications, both layers are coated on transparent conducting electrodes. Transition metal (TM) oxides have been investigated extensively as EC materials. These include cathodically-coloring oxides (e.g. $WO_3$, $MoO_3$) which become colored in the reduced state[26–28], and anodically-coloring oxides (e.g. NiO, $IrO_2$) which become colored in the oxidized state[29,30]. Among these, $WO_3$ is by far the most widely studied EC, with early work dating back to the 1970s[26,31]. Upon insertion of cations (e.g. $H^+$, $Li^+$), the normally colorless $WO_3$ turns deep blue[21]. Despite extensive research, the exact mechanisms underlying coloration remain unclear; they likely involve polaronic absorption and intervalence charge transfer, among other effects[23,29].

Some of the key technical factors limiting the widespread commercial deployment of ECs are switching speed, durability, and lack of spectral control. To overcome these, in particular to enable color-neutral switching across the visible spectrum, metal electrodeposition is a promising route[32]. The performance of an electrodeposition-based EC window, especially its switching speed, can be enhanced by employing a hybrid approach that combines metal electrodeposition on the working electrode with ion insertion in a complementary counter electrode[33]. On the other hand, it is sometimes desirable to selectively tune the optical properties of an EC device in different regions of the electromagnetic spectrum[25]. Spectral selectivity can be achieved by integrating an EC material into a metasurface consisting of subwavelength nanostructures. Ion insertion induces changes in the refractive index and extinction coefficient of the EC, resulting in a modulation of the metasurface's optical response. Successful demonstrations have employed TM oxides[34,35] and polymers[36] as the active EC materials, and show significant potential for energy-efficient, non-volatile optical switching.

While phase transitions are not necessary to induce electrochromic effects, they may offer new opportunities for spectral control. As compared to the simple picture where injected carriers modify the electronic structure of the host via band filling, ion insertion can lead additionally to dramatic optoelectronic changes through the creation of new phases. In the archetypal strongly correlated system $VO_2$ for example, the insertion and removal of $O^{2-}$ or $H^+$ ions can trigger a reversible insulator-metal transition[37,38]. Because the optical properties of the insulating and metallic phases differ primarily in the infrared (IR), this approach can be exploited to create dynamic windows that selectively block the IR part of the solar spectrum, while transmitting visible light[38,39]. Furthermore, in materials that can insert more than one type of ionic species, it is possible to induce ion-selective transformations into different phases. Starting from the brownmillerite $SrCoO_{2.5}$, $O^{2-}$ insertion causes a topotactic phase transition into the perovskite $SrCoO_{3-\delta}$, whereas $H^+$ insertion creates a new phase[40,41] $HSrCoO_{2.5}$. This enables reversible, dual-band EC modulation within the visible and IR regions.

**Electrochemical random-access memory: Modulate electrons**

Emerging computing applications like image recognition, natural language processing and autonomous driving require a computational architecture that can process and store large quantities of data in real-time[42,43]. Present day computers are based on the von Neumann architecture, which is unsuitable for these tasks as it involves moving data between separate units of logic and memory – this is slow and energy-inefficient. The human brain is by far the most complex and energy-efficient computational engine that is known – it performs between $10^{14}$ to $10^{16}$ synaptic operations per second while consuming only ≈20 W of power, or ≈1-100 fJ per operation[44]. To put this in perspective, IBM Watson, the computer that defeated humans in a game of *Jeopardy!* in 2011, consumed ≈85,000 W [45].

Inspired by the architecture of the brain, new computational approaches have been proposed which implement the complexity of neural pathways through arrays of *memristors*[13,46]. The basic building blocks of any such hardware are devices whose electronic conductance can be tuned deterministically. Training a neuromorphic array involves optimizing these conductances (or adjusting the synaptic weights) to minimize error between the output and the solution. By avoiding the need to shuttle data between separate units of logic and memory, this type of computational approach seeks to emulate the superior energy efficiency of the brain. Conventional non-volatile memory technologies such as phase-change and metal-oxide resistive random-access memory have been used previously to build neuromorphic hardware with some success[46–48]. However, the high nonlinearity, asymmetry, and write energies of these devices have posed challenges for high-accuracy, energy-efficient performance. Linearity refers to the extent to which the conductance change induced by a write pulse depends on the initial conductance, while symmetry refers to the difference between the conductance change induced by a positive versus a negative pulse.

Electrochemistry enables reversible control over electronic conductivity, decoupling the "write" (setting conductance) and "read" (measuring conductance) processes using a third terminal. Remarkably, some of the earliest demonstrations of electrochemical-neuron-based learning were made as early as the 1960's, involving the electroplating of graphite leads in liquid cells[49,50]. Here, we limit our discussion to examples where ion *insertion* is the primary mechanism that tunes electronic conductivity. A typical electrochemical random-access memory (ECRAM) device (also referred to as "redox transistor" or "synaptic transistor"), consists of a mixed ion-electron conductor channel into which ions are inserted via an ion-conducting electrolyte under the action of a gate electrode[51,52]. The electronic conductivity of the channel is measured using source and drain electrodes. The electron-blocking nature of the electrolyte enables excellent off-state retention. Note that ECRAMs are distinct from electrolyte-gated (or ionic-liquid-gated) field-effect transistors in which an electric double layer at the electrolyte/channel interface induces carriers in the channel purely through electrostatic effects, i.e. without ion insertion. Interestingly, some studies have shown that the high electric fields in electrolyte-gated transistors can induce ion insertion/removal, suggesting that electrochemical effects may complicate the interpretation of ionic-liquid-gating experiments[37,53,54].

TM oxides are popular candidate materials for ECRAM. Devices have been demonstrated using $Li^+$ insertion in $Li_xCoO_2$ [55], $Li_xWO_3$ [52], $Li_xMoO_3$ [56] and $Li_xTiO_2$ [57], $H^+$ insertion in $H_xWO_3$ [58,59], and $O^{2-}$ insertion in $SmNiO_{3-x}$ [60], $SrCoO_{3-x}$ [61] and $TiO_{2-x}$ [62]. In the simplest model, when cations are inserted into a TM oxide host, the compensating electrons raise the Fermi level, resulting in a modulation of its electronic conductivity. In some cases, more complex effects can occur that are not described by this rigid band model[63,64], as discussed later in this Review. Recent demonstrations of TM oxide-based ECRAM have shown excellent write linearity, symmetry, and low noise, with performance in neural network simulations far exceeding metal-oxide filamentary memristors[55]. Additionally, two-dimensional van der Waals (2D vdW) material based redox transistors offer unique advantages in terms of scalability[65,66]. It is projected that highly-scaled 2D synapses could operate with only attojoules of energy, surpassing biological synapses[67].

Organic electrochemical transistors (OECTs) have received significant attention as memristive elements because they can enable low cost, flexible, and perhaps most interestingly, neuro-compatible devices[11,68]. The most well-studied organic semiconductor for this application is poly(3,4-ethylenedioxythiophene) doped with poly(styrene sulfonate) (PEDOT:PSS) [69]. The backbone of PEDOT is doped with holes, which are compensated by sulphonate ions from PSS. $H^+$ insertion has been used to modulate the electronic conductivity of PEDOT:PSS partially reduced with poly(ethylenimine)[70]. In a significant step towards a neuromorphic system, by combining PEDOT:PSS ECRAM with conventional resistive memory-based selectors, array-level neuromorphic learning has recently been demonstrated[71]. The ability to design and synthesize specialty polymers allows to improve the performance of the ECRAM devices as needed for scaling[72].

**Thermal transistors: Modulate phonons**

Thermal transistors are an emerging class of functional devices whose thermal conductivity ($\kappa$) can be tuned dynamically[73,74]. In nanoelectronics and photonics, thermal dissipation presents a severe bottleneck that limits device performance. Traditional approaches to manage heat are *passive*, i.e. they employ a heat sink with fixed $\kappa$ [75]. For next-generation devices, where heat loads can have complex spatial and temporal profiles, it is essential to develop ways to *actively* route heat at the nanoscale. Additionally, it has been proposed that materials with variable thermal conductance can enable significant HVAC energy savings (≈10 to 40% in U.S. residences) when incorporated into building envelopes, potentially lowering greenhouse gas emissions[76,77].

In general, ion insertion modulates both the electronic and lattice components of $\kappa$. The former can increase upon ion insertion due to the concurrent injection of electronic carriers. However, in many insertion materials, the effects of electronic doping on $\kappa$ are small at room temperature, and dominate only at low temperatures[78–80]. Instead, ion insertion primarily modifies the lattice component, through the introduction of point defects that disrupt the periodic packing of atoms and impede the propagation of thermal vibrations, i.e. phonons. This contrasts with the EC and ECRAM examples discussed above, where property modulation was driven primarily through changes in the electronic structure of the host. The lattice thermal conductivity can be written as: $\kappa \sim Cv^2\tau$, where $v$ is the phonon group velocity, $C$ is the specific heat, and $\tau$ is the phonon relaxation time[14]. The quantities $C$ and $v$ are harmonic properties of the lattice, i.e. they depend on atomic mass and strength of interatomic bonding; $\tau$ is determined by the rate of phonon scattering with other phonons (i.e. anharmonicity), interfaces, and point defects. Electrochemical ion insertion offers a means to dynamically tune each of these properties, and thereby enables significant control over the lattice thermal conductivity.

The mechanisms driving the modulation in lattice $\kappa$ due to ion insertion can be broadly considered under two categories: (1) Point defects (either inserted ions or vacancies) create localized vibrational modes, which act as scattering sites for heat-carrying phonons of the host crystal. This reduces the average $\tau$, and the phonon mean-free-path, $\Lambda = v\tau$. (2) Inserted ions modify the structure of the host material itself, by changing its volume, strength of interatomic bonding, or stabilizing a new phase. This changes the intrinsic phonon dispersion of the host, modifying $C$ and $v$. Phase boundaries can also act as phonon scattering sites.

Early studies of the impact of ion insertion on heat transport date back to the 1980's, where ex situ measurements of graphite intercalation compounds observed a reduction in $\kappa$[78]. While the effects of lattice disorder and strain on $\kappa$ have been extensively studied in the past, only recently have they been exploited to create materials with *dynamically* tunable $\kappa$. The first demonstration of an ion-insertion based thermal transistor was made in the well-known Li-ion battery electrode $Li_xCoO_2$ [79]. Here, Li deinsertion lowered $\kappa$, suggesting an important role played by phonon scattering at Li vacancies. These devices displayed a thermal modulation ratio of ≈1.5x and switched on timescales of a few hours, serving as an important proof-of-concept.

Van der Waals layered materials like graphite and TM dichalcogenides are promising candidates for electrochemical thermal transistors[74,80–83]. This is due to the large coherence lengths of heat-carrying phonons both along the basal planes[84] and across them[85,86], as well as the ease with which ions can be reversibly inserted into the vdW gaps. The long phonon mean-free-paths in the pure material imply that they can scatter strongly with a broad range of defects created during insertion. In some cases, due to competing effects involving phonon softening, scattering, and phase transitions, $\kappa$ may vary non-monotonically with ion concentration[80,83]. For example, in bulk $MoS_2$, Li insertion results in a structural transition from a semiconducting to a metallic phase. Phase coexistence at intermediate Li concentrations can result in stacking disorder along the *c*-axis, causing $\kappa$ to go through a minimum[80]. Furthermore, intrinsic anisotropy in the crystal structure of some vdW crystals can lead to varying degrees of $\kappa$ modulation along different axes[82], offering interesting opportunities for the directional routing of heat.

With the goal of developing thermal transistors with higher switching ratios and faster operation, nanoscale devices are being explored. Using Li insertion into few-nm-thick $MoS_2$, thermal transistors with a large on/off ratio of 10x have been demonstrated[74]. We note that a ten-fold tunability in thermal conductivity is significant considering that the range of $\kappa$ in natural solids spans only 4 to 5 orders of magnitude[87]. This contrasts with the much larger range of electronic conductivities, which spans >30 orders of magnitude. For practical use as a thermal regulator, a ten-fold improvement in the on/off ratio of a thermal transistor represents a potential enhancement of ~ thousand-fold in the lifetime of a temperature-regulated device[74].

The atomic and mesoscale structural phenomena that lead to modulations in $\kappa$ are known to occur in a wide variety of ion-insertion compounds studied extensively by the energy storage community. There is therefore a vast materials space of electrochemically-driven thermal transistors that remains to be explored, such as systems that can host multiple ionic species and display bidirectional tuning[40,88]. From the application standpoint, the key parameters that need improvement are switching speed and lifetime (especially if a phase transition is involved); in fact, this is generally true for all of the switching applications discussed above. Decreasing the length scales of thermal transistors has enabled a reduction in cycle time from several hours[79,82] to a few minutes[74]. Drawing inspiration from the ECRAM community where device response times < 10 ns have been demonstrated[52], further improvements are possible with device optimization.

Finally, we observe that besides demonstrating functional thermal materials using ion insertion, these studies have shown that the thermal conductivity of an intercalating battery electrode depends strongly on its state-of-charge[79]. Even within single electrode particles, inhomogeneities in ion concentration can lead to localized regions of strongly suppressed thermal transport[74]. Such 'hotspots' could be linked to heat generation at the single particle level and have implications for battery degradation and safety. This will become particularly acute as fast charging (< 15 minutes) becomes commonplace. Non-invasive thermal characterization techniques[89] are well-suited to probe these phenomena.

## *Ion insertion for chemical transformation*

Having discussed how electrochemical ion insertion can be used to modulate transport properties for dynamic switching applications, we now turn our focus towards how ion insertion can be used to control chemical transformations.

**Electrocatalysis**

The desire to reduce energy usage in commodity chemical production and to store renewable energy in energy-dense fuels has shifted focus away from using temperature as a means of accelerating reactions towards using electric fields. In these "electrocatalytic" processes, the surface chemistry of the electrode is intimately connected to its ability to mediate current flow (or reaction rate). Importantly, while traditional electrocatalytic materials used precious metals such as Pt, which are expected to be compositionally static during the catalytic reaction, emerging materials based on abundant TMs often are characterized by bulk ion insertion involving species from the electrolyte or reactants. Thus, while catalysis is generally considered to be a surface phenomenon whereby the catalytic reactions only occur on the surface of the catalyst, the catalytic properties of these emerging electrocatalysts can be strongly influenced by their bulk ion-insertion redox reactions. We note that, while ion insertion materials, especially transition metal (oxy)hydroxides, have been widely investigated as electrocatalysts, using ion insertion to improve performance has largely been overlooked. However, recent work suggests if the identity of the inserted ion can be controlled (i.e. alkali metal instead of proton), the activity of the material towards proton-coupled electron transfer reactions can be enhanced by breaking the pH-dependent scaling relationships between the reaction of interest (i.e. oxygen evolution) and the formation of the active site (bulk ion-insertion reaction)[90].

Ion insertion can influence electrocatalytic activity through a variety of different mechanisms. We can categorize the effects of ion insertion into four main groups: structural effects, sequential electrochemical-chemical (E-C') type mechanisms, electronic effects, and ion shuttling/co-catalyst effects, as shown in **Figure 2**. Often more than one of these effects occurs through ion insertion converting an inactive material into an active electrocatalyst, or vice versa.

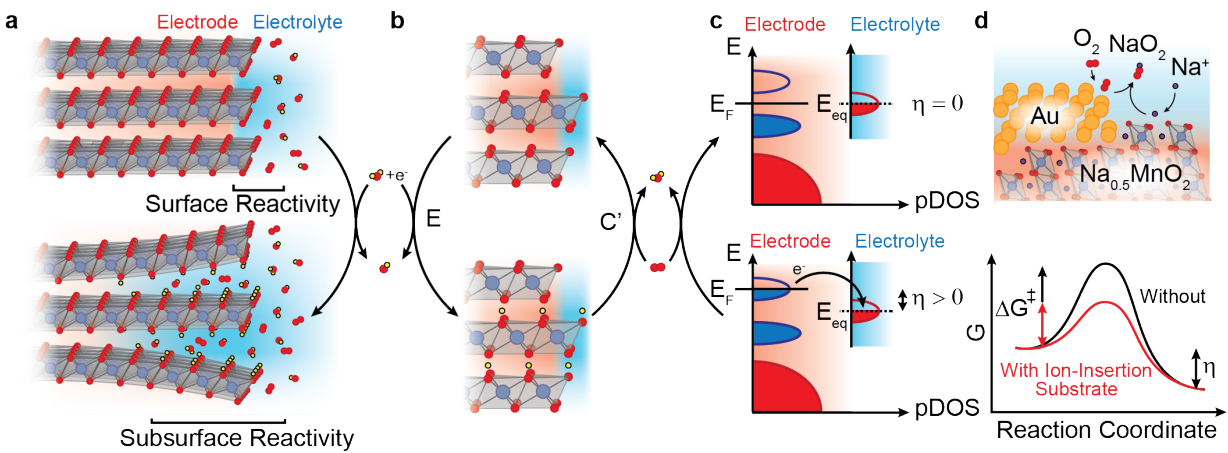

**Figure 2: Ion insertion effects on electrocatalytic electrodes and reactions.** (**a**) Ion insertion may induce structural effects at the mesoscale by expanding the electrochemically active surface area beyond the electrode-electrolyte interface of the host material. (**b**) E-C' type mechanisms whereby a host material incorporates an intermediate reactant of an electrocatalytic reaction through electron-coupled ion insertion (E). This process produces a surface with increased activity that can chemically react with a species of interest through the transfer of both ions and electrons across the electrode-electrolyte interface (C'). (**c**) Ion insertion changes the electron population of the host structure changing both electronic conductivity of the electrode and the thermodynamic electrochemical driving force of the reaction ($\eta = E_F - E_{eq}$, where $E_F$ is the Fermi level of the electrode and $E_{eq}$ is the Fermi level of the reactant in solution). Additional kinetic effects occur based on the stabilization of the $E_{eq}$ of a reactive intermediate with the surface of an electrocatalyst and the overlap of electronic states between the electrode and reactant. (**d**) Co-catalyst or shuttling effects. The kinetic barrier of a reaction can be reduced due to reactant spillover between a non-redox active electrocatalyst and an ion insertion substrate.

While ion insertion is generally assumed to occur topotactically, whereby the host structure maintains its general structure with only minor variations in bonding, it may introduce mechanical strain causing structural transitions that extend from the atomic to the mesoscale. Often, the structural transitions caused by this mechanical strain result in the co-insertion of electrolyte, which may help or hurt the catalytic properties of the electrode. For example, in layered compounds ion insertion increases the vdW gap spacing, which increases the electrochemically active surface area (ECSA) and the resulting catalytic current. In bilayer $MnO_2$, alkali ion insertion increases the electrocatalytic activity for the oxygen evolution reaction (OER), trending with the size of the intercalating ion[91]: $Cs^+ > Rb^+ > K^+ > Na^+ > Li^+$. It has been proposed that the activity increases due to the co-insertion of water and exposure of Mn centers in the interlayer of the material to water or hydroxide ions, which could serve as additional active sites, as demonstrated schematically in **Figure 2a**. In battery materials such as $LiCoO_2$ and $LiCoPO_4$, delithiation during the OER causes dramatic structural changes converting $LiCoO_2$ towards the less active spinel $LiCo_2O_4$ and converting $LiCoPO_4$ towards a more active Co-Pi like structure[92]. In $MoS_2$, lithiation or sodiation exfoliates the $MoS_2$ layers resulting in increased ECSA and improved electrocatalysis towards the hydrogen evolution reaction (HER)[93–96]. In addition, the repetitive insertion and removal of ions during electrochemical cycling, as might occur in the operation of air electrode catalysts of metal-air batteries or regenerative fuel cells, often leads to severe structural degradation and amorphization of the catalyst surface structure[97,98]. Importantly, these transformations are often non-reversible and thus may be seen as extrinsic where ion-insertion is used as a synthetic strategy towards the preparation of new (meta)stable "activated" structures.

As we move our analysis from the meso to the atomic scale, ion insertion can alter the chemical reaction pathways and be used to promote E-C' type electrocatalytic mechanisms[90]. In these mechanisms, an electrochemical step (E) occurs first, whereby electrons are transferred into/out of the external circuit and are coupled to ion insertion into/out of the electrocatalytic material into the electrolyte. A following chemical step (C') occurs where the newly formed phase of the electrocatalyst reacts through a coupled ion-electron transfer with a reactant in the electrolyte, but no net electrons are transferred into/out of the external circuit. In this process, the chemical step reforms the state of the electrocatalyst prior to the "E" step, which is signified with the apostrophe symbol attached to "C". Ultimately, these E-C' processes are governed by the relative reaction rates of the E and C' steps, where the rate of the E step can be increased through controlling the voltage. The C' step is coupled to the E step through reactant/product concentration, as well as by the relative concentration of the active phase at the surface. If the E step is faster than the C' step, then the current will reach a limiting value when the surface concentration of the active phase saturates, i.e. E produces the active phase faster than it is consumed and there is a limit to how much active phase may be produced. This may occur at long times after the entire bulk has converted, or if the reaction rate of E is severely limited by solid-state ion transport in the bulk of the electrode. On the other hand, if the C' step is faster than the E step, then the catalytic reaction should prevent the bulk transformation of the electrode from the ion insertion reaction and the current-voltage behavior of the electrode should follow general electron transfer formalisms, i.e. the reaction appears as only electron transfer with no chemical steps on the phase of the electrocatalyst prior to the "E" ion-coupled reactions. With these constraints in mind, an E-C' mechanism that can be observed implies that the chemical C' step is slower than the ion insertion E step. It is postulated that this mechanism is influential on several TM oxide electrocatalysts in aqueous solutions, where the thermodynamics of the TM-$H_2O$ equilibria governs the proton content of the oxide. For example, as shown schematically in **Figure 2b**, during the oxygen reduction reaction (ORR) on $MnO_2$ based catalysts in alkaline electrolytes, protons are electrochemically inserted into the bulk and surface structure of the electrocatalyst electrochemically converting $MnO_2$ to MnOOH which reacts with $O_2$ according to the following series of reactions:[99,100]

E: $Mn^{4+}O_2 + H_2O + e^- \rightarrow Mn^{3+}OOH + OH^-$   (x4)        [1]

C': $4Mn^{3+}OOH + O_2 \rightarrow 4Mn^{4+}O_2 + 2H_2O$        [2]

The net reaction of the combined steps [1] and [2] is the ORR: $O_2 + 2H_2O + 4e^- \rightarrow 4OH^-$. This process has been shown to occur chemically: bubbling $O_2$ into the electrolyte and MnOOH at open-circuit causes the open-circuit voltage to evolve towards that of $MnO_2$ [99]. The implication being that both electrons and protons are transferred between $MnO_2$ and $O_2$ across the electrode-electrolyte interface without electrons supplied by an external circuit. Analogously, aprotic oxygen reduction in organic media on α-$MnO_2$/R-$MnO_2$ shows similar behavior: Li ions are electrochemically inserted into the $MnO_2$ structure, and can be extracted as $Li_2O$ upon reduction of $O_2$ [101]. Similarly, the OER on Ni-borate and Co-phosphate catalysts involves proton-coupled electron transfer pre-equilibria, which de-intercalate protons from the amorphous catalysts followed by a chemical step that results in the turnover of oxygen and regeneration of the original catalyst[102,103]. In general, E-C' type mechanisms occur when the catalyst serves as a host for an intermediate reactant of the overall catalytic cycle. For example, hydrogen insertion into $WO_3$ can be used to chemically reduce hydrogen peroxide through reaction [3,4] [104,105] or to evolve hydrogen through reactions [3,5] [106–108].

E: $xH^+ + xe^- + WO_3 \rightarrow H_xWO_3$        [3]

C': $2H_xWO_3 + xH_2O_2 \rightarrow 2WO_3 + 2xH_2O$        [4]

C': $2H_xWO_3 \rightarrow xH_2 + 2WO_3$        [5]

While the above reactions generally involve positively charged protons as the inserted ion, anions can also play a role in oxygen defective structures. For example, it has been shown through differential electrochemical mass spectrometry that hydroxide insertion into covalent perovskite oxides, for example $SrCoO_{3-x}$, activates lattice oxygen into the reaction mechanism for the OER through the following reactions [109–112].

E: $SrCoO_{3-x} + 2x\ OH^- \rightarrow SrCoO_3 + xH_2O + xe^-$        [6]

C': $SrCoO_3 \rightarrow SrCoO_{3-x} + x/2\ O_2$        [7]

We note that the mechanisms of reactions [6] and [7] are not limited to proton conducting liquid electrolytes and are well known for high temperature hydrogen or oxygen evolving/reducing electrocatalytic reactions catalyzed by similar perovskite oxide structures. These reactions, which occur at the electrode/gas interface and form the basis for the operation of solid oxide fuel cells, are influenced by the voltage controlled stoichiometry of either protons or oxide ions that are electrochemically inserted from solid state protonic or oxide ion conducting electrolytes[113].

The next way ion insertion influences electrocatalysis, and certainly one of the clearest to directly observe, is via the modulation of the electronic structure of the catalyst. Often this serves two purposes—it changes the electron filling of the active site, converting it to a more active oxidation state, and it modulates the carrier density of the electrocatalyst resulting in a more conductive compound. For example, the lithiation of $MoS_2$ has been used to tune its activity for the HER. Lithiation of $MoS_2$ converts the atomic structure of the catalyst from the semiconducting 2H phase to the metallic 1T phase and reduces the $Mo^{4+}$ centers to lower oxidation state $Mo^{3+/4+}$. These effects have been demonstrated to dramatically increase the activity of $MoS_2$ for the HER approaching the performance of Pt [93–96]. Interestingly, in $MoS_2$ and other TM sulfides, the 1T structure appears at least metastable over long term hydrogen generation, despite re-delithiation during operation (which results in the same stoichiometry as the less active 2H phase)[94,114]. The creation of metastable phases with enhanced properties through ion insertion in generally a useful materials design approach[115]. As mentioned above and shown schematically in **Figure 2c**, proton ion insertion into $MnO_2$ modulates the oxidation state of Mn towards a mixed $Mn^{3+/4+}$, which increases its activity towards the ORR [99,100]. Similarly, oxygen insertion into covalent perovskite oxides lowers the Fermi level of the oxide into the non-bonding O-$2p$ states, resulting in holes on the ligand oxygen[109,116]. From density-functional theory (DFT) based calculation, these oxidized oxygen ligands are then activated and incorporated as reactants in the mechanism for oxygen gas generation [109–112]. In the layered oxides $Li_a(Ni_xCo_yMn_z)O_2$, $LiCoO_2$, $LiCo_{0.5}Ni_{0.5}O_2$, $LiCo_{0.5}Fe_{0.5}O_2$, and $LiNi_{0.33}Co_{0.33}Fe_{0.33}O_2$, lithium removal results in improved carrier concentration and electronic conductivity as well as an increased oxidation state for Co (or potentially ligand holes on oxygen). This results in changing charge-transfer characteristics and intermediate binding strengths generating better bifunctional activity towards the OER and ORR [117–119].

Lastly, ion insertion materials can be used as co-catalysts where the ion insertion properties can be used to enhance the activity of a primary electrocatalyst. One application of this phenomenon is using battery materials as epitaxial supports for Pt nanoparticle ORR catalyst to introduce strain into the Pt lattice. Tuning lithium stoichiometry from $Li_{0.5}CoO_2$ to $LiCoO_2$ was found to introduce compressive strain and boost Pt ORR activity[120]. Ion-insertion materials may also serve as a shuttle or ion-conducting electrolyte for intermediates involved in the overall catalytic turnover. In this case, the local ionic concentration of an intermediate can be substantially increased in the solid state compared to the liquid electrolyte. For example, as shown schematically in **Figure 2d**, a $MnO_2$/Au composite catalyst was investigated for the ORR in Na-air batteries. Using environmental transmission electron microscopy, it was found the reduction of oxygen to superoxide occurred on Au and Na electrochemically inserted into $MnO_2$ to form a $Na_{0.5}MnO_2$ phase.

The Na in $Na_{0.5}MnO_2$ was shuttled to the triple-interface between $MnO_2/Au/O_2$ to form the $NaO_2$ product[121]. Hydrogen spillover from Pt onto $WO_x$ during the methanol oxidation[122] has been confirmed through in situ electrochromism measurements and increases the electrocatalytic activity of the composite catalyst through a combination of reactions [3] and [8].

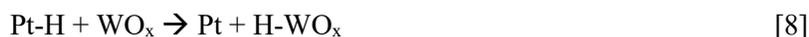

Pt-H + $WO_x$ → Pt + H-$WO_x$                              [8]

Similarly, hydrogen spillover has been used to enhance methanol oxidation by coupling $MnO_2$ with Ru (reaction [1]) [123] and in "bimetallic" Pt-Ru alloys where the Ru is a hydrous oxide, $RuO_xH_y$, that can incorporate protons during the reaction [124,125]. Thus, coupling non-redox-active electrocatalysts with supports that can be reduced through ion insertion is an effective strategy to reducing the kinetic barriers in a variety of electrocatalytic oxidation reactions.

**Water desalination and ionic separations**

There is growing interest in the use of ion insertion materials for water desalination and purification via capacitive deionization (CDI) and ion-selective electro-sorption[126,127]. Prussian Blue analogues (PBAs), such as nickel hexacyanoferrate (NiHCF, $Ni_2Fe(CN)_6$), have been used for decades to separate cations from electrolytic solutions[128]. Recently, these insertion compounds have been applied in CDI systems for desalination[129–134], focusing on the removal of sodium from brackish water. Manganese, copper and zinc hexanoferrates have also been considered, and the former two PBAs are stable alternatives for sodium and calcium insertion[135]. Energy-storage insertion materials, such as sodium manganese oxide (NMO, $Na_2Mn_5O_{10}$)[136–138] and titanium disulfide ($TiS_2$)[139], have also been introduced to this field as a means of both sodium desalination and extraction of energy from water salinity differences.

Ion-insertion materials have higher storage capacity than traditional carbon aerogel electrodes used in CDI, since they store ions in bulk volumes rather than in surface double layers. If ion-insertion reactions and solid diffusion are fast[140], then these systems can be designed to achieve comparable or better energy efficiency compared to carbon-based CDI[141,142], and a recent technology comparison shows that the PBA NiHCF is emerging as one of the most promising sodium extraction materials in this class[143]. An important means to achieve high efficiency is to recover some of the undissipated free energy extracted from composition differences during portions of the cycle when the system produces power [141–144], thus further blurring the distinction among hybrid systems acting as both a rechargeable battery and an electro-sorption system.

In general, electrochemical processes of water desalination become efficient at low salinities, typically well below that of seawater. At high salinities it costs more energy to remove the ions from solution than to directly remove the water by reverse osmosis[145], a mature and commercially available technology. As such, ion-insertion materials hold the most promise for water desalination and purification at low salinity, especially when electrochemical selectivity is desirable. In particular, an attractive feature of insertion materials is their high selectivity for certain target ions, such as lithium. For example, lithium ions can be extracted from sodium or potassium chloride solutions by $Li_xFePO_4$ electrodes with several hundred-fold selectivity[146–148], opening possible applications in lithium harvesting from brines or seawater[149,150]. Insertion-based water desalination or separation systems can also be enhanced by the use of ion-exchange membranes[151,152], which block the rejection of co-ions as counter-ions are adsorbed by a polarized electrode. The engineering tradeoffs in energy efficiency, water recovery, flow rate, and selectivity in these systems can be guided by mathematical models of insertion-based desalination based on porous-electrode theories[140,142]. Such models are analogous to those developed for batteries[153], thus highlighting the universal physics of ion insertion reviewed here.

## Unifying principles

Having established the diverse use cases for ion-insertion solids for dynamic switching and chemical transformation, we now turn our attention to fundamental chemical, physical, and structural concepts that unify ion-insertion solids. In this section, we span length scales from small to large, starting with atomic-scale point defects, interfaces, and finally ending with mesoscale phase separation.

### *Point defects in the host material*

Atomistically, ion insertion is mediated by the formation and annihilation of point defects involving the ionic guest species (generally vacancies or interstitials) and the concomitant electronic compensation (redox) of the host[154]. Often, property changes during ion insertion can be explained solely through consideration of these point defects. For instance, the four orders of magnitude increase in the conductivity of $Li_xCoO_2$ upon delithiation is readily explained by the partial depopulation of Co 3$d$–O 2$p$ $t_{2g}$ electronic states[155,156]. The change in oxidation state and resulting conductivity change with Li content is also associated with modulations in the interslab spacing in layered materials[157]. However, in more complex situations, point defects in the host material, which here we classify as those involving species other than the inserted ion or its associated redox couple, must be accounted for to rationalize and control the properties of ion insertion materials. Here, we review salient examples demonstrating the important yet often-overlooked role of point defects in the host material. We focus mainly on lithium TM oxide battery positive electrodes with the goal of establishing general principles that can inform improvement strategies for emerging ion-insertion applications.

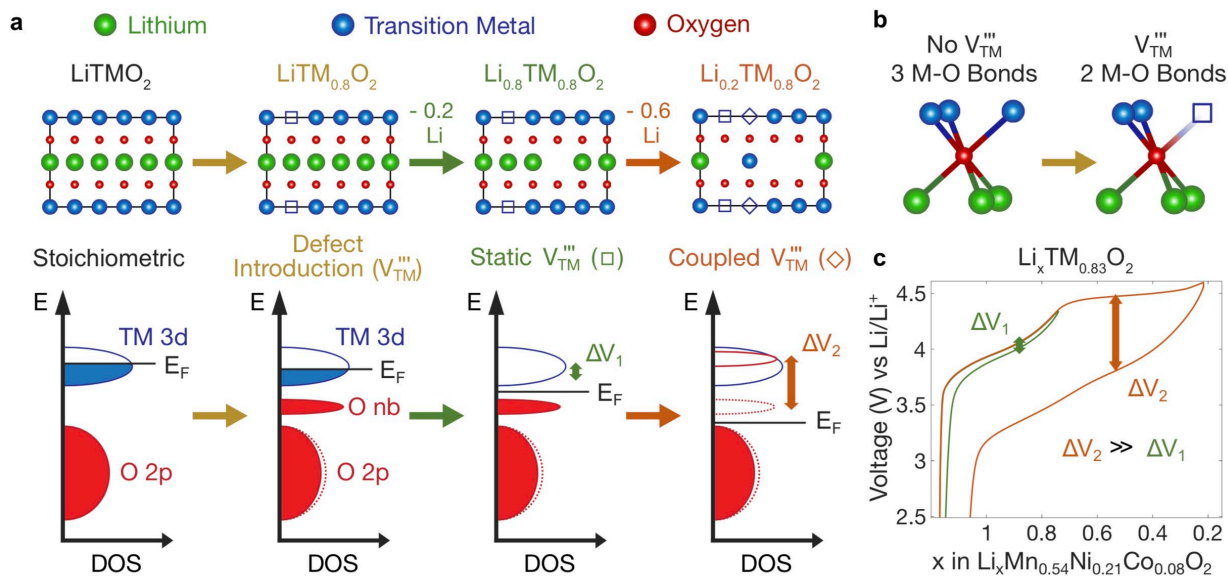

**Figure 3: The role of point defects in the host.** (**a**) Point defects in the host material can be broadly classified into two categories. The first category, here termed 'static defects', includes defects which do not change in concentration during ion insertion/removal. These defects are typically introduced during material synthesis. For $Li_xTM_{1-y}O_2$, if the Li content is kept above x ≈ 0.8, the $V'''_{TM}$ concentration generally remains constant and $V'''_{TM}$ can thus be considered 'static defects'. However, upon deep delithiation, the concentration of $V'''_{TM}$ becomes coupled to that of the inserting ion, causing $V'''_{TM}$ to become 'coupled defects'. Coupled defects can modify the band structure during ion insertion/removal and can even change the relative ordering of electronic states. (**b**) Defects in ion-insertion materials can modify the local bonding environment. Here, $V'''_{TM}$ (or $Li''_{TM}$) decrease the number of covalent bonding partners for oxygen[162,165]. (**c**) Voltage profiles of a Li-excess layered oxide taken from literature; panel adapted with permission from Ref. [166]. In the presence of only static defects, the rigid band model can usually be applied, and thus the voltage hysteresis is often minimal ($\Delta V_1$). Conversely, coupled defects can significantly modify the band structure during ion-insertion (**Figure 3a**), which can lead to a much larger voltage hysteresis ($\Delta V_2$).

In the simplest case, the concentration of a given point defect in the host material is taken as static during ion insertion (**Figure 3a**) and thus does not change during device operation. This "quenched" scenario reflects the slow kinetics of defect equilibration rather than thermodynamics, since mass action relationships generally link all point defects to the inserted ion and/or the redox couple. Even in this situation, the behavior of point defects in ion-insertion solids deviates from that in classical semiconductors for two primary reasons. First, disordered point defects can be present in ion-insertion solids in several to tens of at. %, in contrast to the ppm levels typical of doped semiconductors. Second, point defects in ion-insertion solids, unlike most donor or acceptor dopants in classical semiconductors, can significantly alter the local bonding environment (e.g. number of covalent bonding partners, **Figure 3b**). Due to both of these differences, point defects in the host can considerably alter the electronic structure by shifting the Fermi level across entire bands or creating new bands altogether. One notable example is the substitution of up to one-third of the oxygen anions in $Li_xTMO_3$ ($0 < x < 2$) with fluorine, $F_O^{\cdot}$ (Kröger–Vink notation[15]). This donor defect raises the Fermi level, in some cases across multiple redox couples, at a given Li concentration[158,159,160]. This strategy was utilized to activate the $Mn^{2+/4+}$ redox couple in $Li_xTMO_3$ materials during Li insertion, where Mn otherwise exists as $Mn^{4+}$ at all Li contents[161]. Another important example is TM vacancies, $V_{TM}'''$, in $Li_xTM_{1-y}O_2$ ($0 < x < 1+y$) layered oxides. Here, these vacancies (or $Li_{TM}''$) create a new redox couple in the form of a non-bonding O $2p$ band (**Figure 3a,b**)[162,163]. The relatively labile oxygen electrons originating in this band can be extracted at reasonable voltages, allowing for higher battery electrode capacities at ~ 4.5 V vs. $Li/Li^+$ on the initial delithiation (oxidation)[164,165].

In some materials, point defects are not quenched and respond to ion insertions through multiple defect equilibria. In the aforementioned $Li_xTM_{1-y}O_2$ layered oxides, the $V_{TM}'''$ concentration generally remains constant if the Li content is kept relatively high ($x > ~ 0.8$)[166]. However, upon more significant delithiation, a large voltage hysteresis between delithiation and re-lithiation is typically observed (**Figure 3c**), pointing to a dynamic structure change that cannot be explained by the rigid band model. This unusually large voltage hysteresis, which persists even at vanishing ion insertion rates[166], has recently been linked to an increase in the $V_{TM}'''$ concentration through the formation of TM anti-site/vacancy defect pairs[167] ($TM_{Li}^{\cdot\cdot}$ and $V_{TM}'''$, so-called TM migration). Upon deep delithiation, the $V_{TM}'''$ concentration therefore becomes coupled to the ion insertion process. DFT calculations, which are frequently utilized for understanding the effect of defect species on the local electronic structure[168], indicate that the increase in $V_{TM}'''$ concentration causes a significant change in the local electronic structure due to metal–oxygen decoordination[166,167,169] (i.e. the breaking of M–O bonds, **Figure 3a**). In this case, the oxidized form of the redox couple, $O_O^{\cdot}$, is located adjacent to one or more $V_{TM}'''$, with their proximity being driven by the minimization of interatomic strain energy associated with localizing an electron hole. For this reason, the two defects are considered as a complex[15], $(O_O^{\cdot}V_{TM}''')''$. Although the exact nature of the defect complex is debated and may vary among different materials, the overall defect reaction occurring at large extents of delithiation can be written in a general form as:

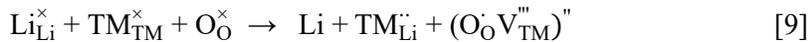

$$Li_{Li}^{\times} + TM_{TM}^{\times} + O_O^{\times} \rightarrow Li + TM_{Li}^{\cdot\cdot} + (O_O^{\cdot}V_{TM}''')'' \qquad [9]$$

While TM migration is coupled to the Li content upon deep delithiation, this process is not fully reversible under typical cycling conditions. This imperfect reversibility results in intra-cycle voltage hysteresis (**Figure 3c**) and a buildup of $TM_{Li}^{\cdot\cdot}$ and $V_{TM}'''$ over extended cycling. Mitigating electrochemical hysteresis by preventing the formation of coupled defects at low Li content is currently an active area of research[170].

Point defects in the host can also affect ionic transport in ion insertion materials, most commonly by expanding or contracting the lattice. For instance, $TM_{Li}^{..}$ defects typically contract the Li layer in Li-ion layered oxides, which can raise the activation barriers for Li$^+$ transport by hundreds of meV[171]. Conversely, oxygen vacancies, $V_O^{..}$, in Li$_x$MoO$_{3-\delta}$ expand the vdW gap and nearly eliminate diffusion limitations even at cyclic voltammetry scan rates as large as 100 mV/s[172]. Point defects can also open new ion migration pathways. This case can be found in γ-MnO$_2$, which stores energy via the fast insertion/removal of protons (or alkali metal cations)[173,174,175]. In γ-MnO$_2$, Mn vacancies, $V_{Mn}^{''''}$, can be introduced which are surrounded by four protons residing on oxygen sites to form the defect $OH_O^{.}$. Inserted protons typically move along 1-D tunnels, but when encountering a Mn vacancy, can use the compensating $OH_O^{.}$ defects to cross over to an adjacent tunnel via a chain-type mechanism[176,177]. This vacancy-enabled migration pathway improves proton transport and high-rate electrochemical performance[178]. Interestingly, defect-defect interactions between mobile ions can also play an important role in ionic transport, a point which we will return to later.

One complication in studying point defects is that they can be spatially heterogeneous due to sluggish transport and/or surface effects. For example, in Li-ion layered oxides, contact with organic electrolytes alters the atomic structure near the particle surface[179]. This surface region, generally ~ 2–10 nm in thickness[180,181], typically contains higher concentrations of $TM_{TM}^{'}$ and $TM_{Li}^{..}$ point defects than the bulk, and can form a crystallographically distinct surface phase (e.g. spinel or rocksalt)[182,183,184]. Even in the particle bulk, the type and concentration of relevant defects can vary at the nanoscale. For instance, electron microscopy studies have shown that TMs which migrate to the Li layer can exist predominately in octahedral ($TM_{Li}^{..}$) or tetrahedral ($TM_i^{...}$) sites in different regions of the same particle[185,186]. These heterogeneities highlight the necessity to utilize complementary local and bulk-averaged probes to holistically explain the behavior of complex materials. In other words, a purely local view may misinterpret the underlying mechanisms, while a purely global view may overlook the influence of inhomogeneities.

On the other hand, spatial heterogeneity in defect concentrations can be exploited for practical benefits. One example is suppressing the formation of surface oxygen vacancies, $V_O^{..}$, which occurs at large extents of delithiation (oxidation) in most Li$_x$TM$_{1-y}$O$_{2-z}$ materials. Increasing the TM stoichiometry (i.e., lowering the $Li_{TM}^{''}$ concentration) significantly mitigates oxygen release but also compromises the larger gravimetric capacity afforded by the higher Li content[187]. By creating particles with a compositional gradient in which the surface has a significantly higher TM concentration than the bulk, oxygen release was virtually eliminated while maintaining the sizable capacity of high Li content materials[188]. This example, which has been applied in other variations as well[189–192], illustrates the manipulation of heterogeneity in the host's defect landscape to improve device performance. Optimizing point defects in the host structure is expected to play a similarly crucial role in emerging ion-insertion applications.

## Interfaces

**Coupled ion-electron transfer**

The microscopic mechanisms involved in electrochemical ion insertion reactions at the electrode/electrolyte interface are complex as they depend intimately on the properties of the host material, the redox couple, the intercalated ions, and the electrolyte. In many applications, such as Li-ion batteries, it is assumed that solid-state diffusion is the slowest process, but it is becoming increasingly recognized that ion insertion is likely the true rate-limiting step for a wide range of (dis)charging conditions[193]. Moreover, recent research has shown that ion-insertion kinetics strongly influence non-equilibrium thermodynamics[153,194], including instabilities and patterns formed by solid-state phase separation[195]. As such, improved understanding of the fundamental mechanisms of ion insertion may open new opportunities for interfacial engineering to improve rate capability and electrochemical stability in batteries and other applications.

**Figure 4a** shows a schematic of the insertion process. Initially, the ions participating in the insertion reaction are located in the electrolyte, which can be either solid or liquid. During insertion, the ion is transferred across the electrode/electrolyte interface, leaving the electrolyte and entering the host material. At the same time, an electron from the electron donor (for cation insertion), which can be either the host material, the surface, or a coating, reduces a solid-state redox couple in the host located near the site where the intercalated ion is inserted. An electron-ion pair is formed when the process is completed, where in many cases the electron is localized in the host material (polaron)[196]. This process involves the transfer of both ions and electrons, which can be a coupled or a sequential process depending on the insertion driving force (overpotential)[194,197].

It is common practice to describe the rate of electrochemical ion insertion using the phenomenological Butler-Volmer (BV) equation[198]. In this framework, the microscopic details of charge transfer are not explicitly specified, although it is usually thought to describe classical ion transfer (IT) biased by the interfacial electric field. It was recently proposed that electron transfer (ET) may instead be the rate-limiting step during ion insertion in the case of the $Fe^{2+}/Fe^{3+}$ redox reaction for $Li_xFePO_4$[199], and Marcus-Hush-Chidsey ET theory was incorporated into Li-ion battery models as an alternative to BV kinetics, thus opening the possibility of reaction-limited current (in addition to diffusion-limited current)[153]. Indeed, at the completion of ion insertion, in many cases a localized electron state, e.g. a reduced transition metal site, is created that can be modeled using polaron formation theory[200,201], which is similar to Marcus electron transfer theory in liquid electrolytes[202]. In all-solid-state systems, the Debye length can be similar (and sometimes shorter) than in liquid electrolytes owing to the high concentration of mobile carriers, but strain accommodation is a major consideration. In both liquid and solid environments, the connection between IT and ET during ion intercalation has only just begun to be understood.

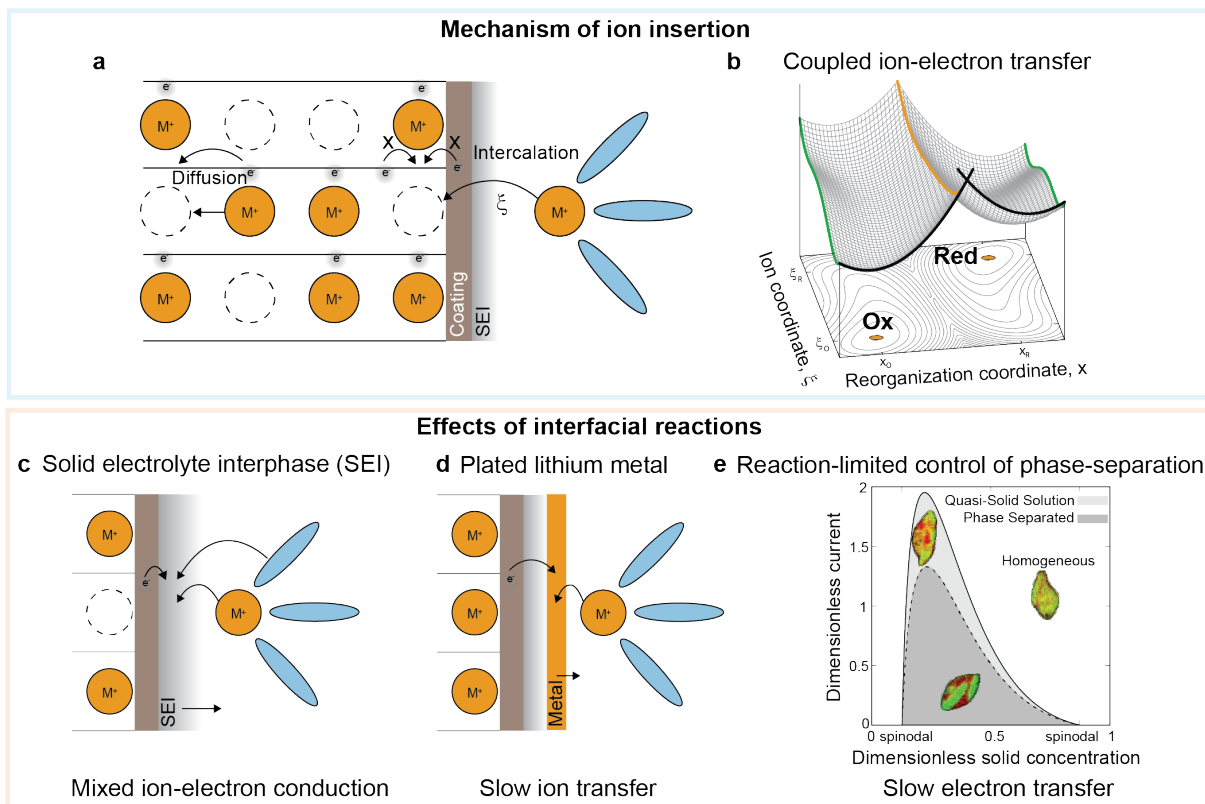

**Figure 4: The microscopic picture of insertion** (**a-b**) involves at least three processes: desolvation of an ion in an electrolyte solution, transfer of the ion into the host crystal, and the transfer of an electron from the host crystal or a surface coating. Either the ion transfer or the electron transfer can be rate limiting, and will depend on the properties of the ion, the electrolyte, and the structure and electrical conductivity of the host crystal. (**c**) If the ion reacts with the electrolyte and an electron from the host, the ion will not be inserted but will instead form a layer of solid electrolyte interphase. (**d**) Alternatively, when there are no available sites for insertion at the surface of the host crystal, the ion can be reduced at the interface to form a layer of plated metal. (**e**) In nanocrystals, nonlinearity in the interfacial reaction rate with respect to composition can either enhance or suppress the phase-separation process, and there is a critical applied current above which phase-separation is suppressed. The inset images show $Li_xFePO_4$ crystals in the phase-separated, quasi-solid solution, and homogeneous regimes. Images in panel (e) adapted with permission from Ref. [233].

Coupled ion-electron theory (CIET) was recently introduced to model this concerted nature of ion insertion processes[197]. CIET considers the short-range electrostatic interactions of the formed ion-polaron pairs, and argues that their formation is followed through a coupled pathway only, where both the ion transfer and the reduction of the host material occur simultaneously. The physical picture of CIET can be represented in the multidimensional excess energy landscape shown in **Figure 4b**, where the reaction coordinates are the polarization of the redox environment $x$ and the ionic coordinate $\xi$, which is equivalent to the distance of the ion from its inserted state. The redox states have a parabolic dependence in terms of the polarization coordinate $x$, as in the classical Marcus theory[203,204]. In the ionic coordinate, however, the functional dependence of the excess energy landscape is much more complex than the ET counterpart, because the ion interacts electrostatically with both its environment and the transferred electron.

As in classical Marcus theory[202], when the redox states intersect they share a common electronic energy level that allows non-adiabatic electron transfer to occur through tunneling. In CIET, the intersection of the two states corresponds to a continuous line in the energy landscape, as depicted with the orange curve in **Figure 4b**. While electron transfer can occur anywhere the reaction complex lies along the intersecting parabolas, the energy required for only electron transfer is prohibitively large as the final state would not satisfy electroneutrality. The same is also true for ion transfer alone. Therefore, CIET predicts that there is a single point across the intersecting line of the parabolas where the transition state barrier is minimum for transferring both the electron and the ion.

The non-adiabatic nature of the electron transfer step during CIET results in non-trivial non-equilibrium response of the system under applied overpotential[205]. In particular, when an energy difference between the redox states is applied, the energy landscape in the polarization coordinate remains parabolic, which has significant consequences for the generated current. In classical charge transfer models, increasing the energy difference between the redox states results in monotonic increase of the resulting current[153,194,206]. This is not the case in electron transfer models. More specifically, for single state electron transfer there is a critical overpotential value – equal to the reorganization energy of the electron acceptor environment[202] – where the electrochemical reaction becomes barrier-less. Larger overpotentials result in the non-intuitive phenomenon of slowing reaction rates, and thus the region of overpotential values for which the current decreases is called the inverted region[202]. This behavior is included in coupled ion-electron transfer, and is responsible for the prediction of reaction-limited currents during ion insertion.

The coupling between ion and electron transfer in ion insertion leads to novel observations that would not be possible in the classical charge transfer picture. For example, the non-monotonic (with respect to composition) and limiting current behavior in ion insertion processes influence the non-equilibrium phase stability of the system of interest[195], leading either to stabilization of a thermodynamically unstable solution, like $Li_xFePO_4$[207] (**Figure 4e**), or destabilization of a stable one, such as $LiNi_{1/3}Mn_{1/3}Co_{1/3}O_2$[208]. Moreover, knowledge of the microscopic picture of ion insertion paves the way for interfacial engineering, e.g. to alter the dielectric properties or charge of the electrolyte/electrode interface, and thus tune electrochemical performance at the macroscopic level. In particular, CIET indicates that when the static and optical dielectric constants are close to each other, the reorganization energy, and thus the electron transfer barrier becomes negligible. Furthermore, the developed theory provides a link between ion-insertion rates and the electronic properties of the electron donor, which in most cases is the insertion material itself. Although much remains to be understood, the rich physics of ion-insertion reactions are beginning to be rationalized by microscopic models.

**Effects of interfacial reactions**

The microscopic nature of ion insertion at interfaces poses dramatically different consequences for different applications. Whereas promoting non-insertion reactions at insertion surfaces is desirable for electrocatalytic applications as previously discussed, for energy storage these are undesirable parasitic reactions that lead to degradation. This section discusses ways in which insertion can go wrong for battery electrodes, in the form of solid-electrolyte interphase (SEI) growth and lithium plating at the interface (see **Figure 4c,d**). Later, we will discuss the important role that reaction kinetics at interfaces play on the stability of phase-separating processes in reaction-limited nanoparticles (**Figure 4e**).

Lithium ions may react irreversibly with the electrolyte rather than intercalate, leading to the formation of SEI that coats the insertion surface. SEI is a mixed conductor of both ions and electrons[209] and tends to grow at high currents during deinsertion[210]. A stable compact layer of SEI protects the surface of the host material from reacting with the electrolyte, but the growth of unstable, thick SEI layers consumes lithium and can significantly increase resistance, leading to degradation. SEI grows primarily on graphite anodes and is highly dependent on electrolyte composition[211–213], although SEI growth has also been reported on positive electrode materials[214–216]. SEI growth is typically mitigated through electrolyte engineering and carefully designed battery formation cycles[217]. If the electrode potential falls below the standard potential of the zero-valent species (e.g. $Li^0$) and there are no empty surface sites, there will be a competing energetic driving force to reduce ions (in the case of cations) onto the surface of the host material rather than to insert. When this occurs, electrons are transferred from the host material to form neutral metal (e.g., Li) on the surface. For Li, this plating process is reversible when it occurs to a small degree, but it becomes irreversible when Li mechanically detaches from the surface or reacts with electrolyte to produce SEI.

*Phase separation and thermodynamics*

The influence of phase separation on ion insertion (and vice versa) has been a recent research focus, particularly for battery electrodes, with lithium insertion into $Li_xC_6$, $Li_xCoO_2$, and $Li_xFePO_4$ as the most familiar examples[218–220]. As a means to control microstructure, phase separation is also applicable to phonon scattering at phase boundaries in thermal transistors and the surface phases of electrocatalytic materials. Although ion insertion occurs on the length and time scales of atoms and electrons, phase stability is governed by bulk thermodynamics at macroscopic scales; the driving force for ion-insertion materials to phase-separate is chemical in nature, a positive free energy of mixing. Microscopic and macroscopic physics therefore meet at the insertion interface. **Figure 5** illustrates the range of length scales over which phase separation has been observed in insertion materials, spanning from the scale of atoms to the scale of devices. A recent theme in the study of phase separation in insertion materials has been the application of classical concepts from thermodynamics. This section focuses on the important role that free energy, F, plays in determining the equilibrium state of phases in insertion materials across length scales.

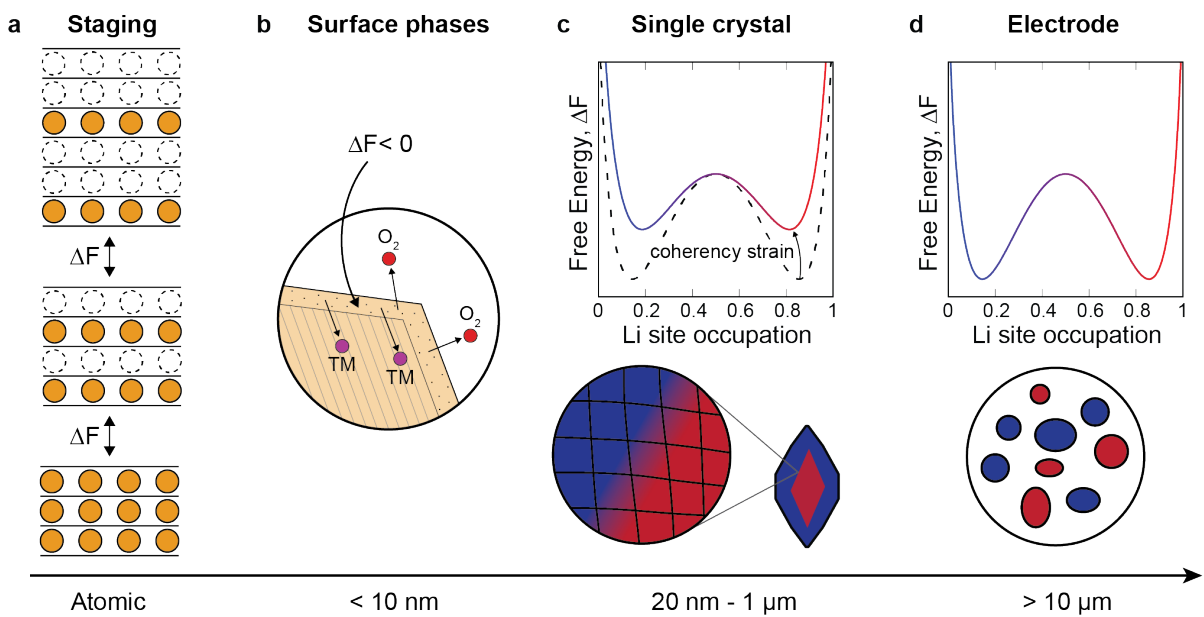

**Figure 5: Phase-separation in insertion materials across length scales, ranging from the atomic scale to the device scale.** The driving force for phase-separation in all cases is a decrease in free energy, often resulting in two-phase coexistence at equilibrium. (**a**) Staging is an ordering process which occurs at the atomic scale. (**b**) Structurally different phases can form on the surface of insertion materials at the nanometer scale. At larger length scales, phases may (**c**) coexist in a single crystal with the formation of phase boundaries, or (**d**) alternatively can form a mosaic pattern of single-phases in a polycrystalline system. All the phenomena pictured are reversible with the exception of surface phase formation, which involves irreversible structural changes and is often accompanied by the loss of oxygen and the migration of TM atoms into the host crystal.

The phenomenon of staging can be thought of as a phase-equilibrium process that occurs at the length scale of atoms. Staging is an ordering process where the inserted ions (or vacancies) in the host lattice are not randomly distributed, but instead preferentially segregate into a periodic arrangement. An example of lithium staging in graphite is illustrated in **Figure 5a**, where every atomic sited is filled when the host is at 100% Li insertion, every other plane is filled at 50% insertion, and every third plane at 33% insertion. These stages correspond to free energy minima, and the conversion of one stage to another during insertion involves traversing a two phase-region, which produces a voltage plateau. Each stage in graphite has a different color observable with optical microscopy[221], which has enabled tracking the evolution of stages at both the particle and electrode scales[222,223]. Lithium/vacancy ordering has also been reported in other battery materials $Li_xCoO_2$[224], $Li_xNiO_2$[225], and $Li_xFePO_4$[226], and layered ordering of oxygen vacancies has been observed in perovskite oxides during electrocatalysis[109].

At a larger length scale ranging from a few nanometers to a few tens of nanometers, it is common for a thin layer of a new phase to form at the insertion surface which is structurally[179,180] or electronically[227] different than the bulk (**Figure 5b**). The structural phase transformation process is driven by a negative formation enthalpy which develops during insertion/deinsertion[179,180]. Although these phases are small in extent, they have a significant influence on ion insertion because they form directly at the insertion surface. The inserted ion and electron must interact with this phase before reaching the bulk, placing importance on its ionic and electronic transport properties as well as its defect concentration. Surface phases can also alter the optical properties of materials. For example in the electrochromic NiO, a surface phase of NiOOH rather than the bulk is responsible for coloration upon the extraction of protons[29].

In Li-ion layered oxides, the surface phase is typically a spinel or rock salt structure that tends to inhibit lithium insertion, leading to capacity loss and impedance increase. The phase begins to form at high voltage after roughly 60% of the lithium has been removed[179], and grows over the course of many electrochemical insertion cycles. The thickness of the phase may also depend on the applied current[182]. Typically, This structural phase change is often accompanied by oxygen loss at the surface[181,182], TM migration from the surface into the bulk[181], Li-TM migration into Li sites[181,184], and changes in the oxidation states of TMs[179,183,228], all of which make the phase transition irreversible. This contrasts with the other phase change processes discussed in this section, which are reversible. Consequently, a strategy to mitigate the formation of surface phases involves the application of engineered surface coatings[229,230].

Many insertion materials experience an energetic driving force to phase-separate into enriched and depleted regions at intermediate extent of insertion rather than to evolve toward a homogenous state. The most common phase-separating host materials are $Li_xCoO_2$, $Li_xFePO_4$, and $Li_xC_6$, although phase-separation has also been reported in other layered oxides[231]. Phase evolution in graphite crystals has been imaged optically[222,232], and advanced experimental techniques have enabled high-resolution imaging of phases evolving in single crystals of $Li_xFePO_4$[233–237] and $Li_xCoO_2$[238]. As illustrated in **Figure 5c**, the composition-dependent free energy of the intercalating system typically exhibits minima separated by a barrier. When phase separation occurs within a single crystal (e.g., primary particle), a phase boundary with an interfacial energy will form between the two phases. The two phases will typically have different lattice constants, and as a result will exert mechanical forces on each other. This additional coherency strain energy modifies the free energy curve, lowering the free energy barrier between phases and making phase separation less energetically favorable[239]. However, changes in the unit cell volume during ion insertion enhances degradation in both single-phase and two-phase materials[240].

The orientation of phase boundaries can be affected either by the equilibrium properties of the system, or by the non-equilibrium ion insertion pathway. Coherency strain, for example, influences the orientation of phase boundaries[241], resulting in the tendency for alignment along elastically-preferred orientations[239]. Likewise, the crystal size as well as the surface properties will affect both the morphology of the formed interfaces as well as the conditions under which they form[241,242]. There are cases, however, where strongly anisotropic transport properties result in morphologies different from the thermodynamically predicted ones[243]. This behavior arises from the continuous input of energy during non-equilibrium ion insertion. When the particle size is larger than the interfacial thickness, there is a competition between coarsening and ion insertion that determines the final phase morphologies observed during ion intercalation[193]. When the total insertion rate is slower than the diffusion timescale of the slowest forming phase, intercalation wave structures[207,244] are observed. At the other limit where the coarsening of the slowest phase is the rate-limiting step, shrinking-core structures[245] evolve. In graphite, this leads to poor utilization of the active material and is responsible for the onset of Li plating. Applied electric fields also contribute to phase stability and the morphology of the formed phases. For example, insertion materials that undergo metal-to-insulator transitions[243,246] have been shown to phase-separate and form filament-like structures[247] when large voltage drops are applied across the system.

Phase-separation behavior changes significantly when the size of the particle approaches the phase boundary width. Below a critical particle size, phase-separation will be completely suppressed[239,248]. Above the critical particle size, interfacial reaction kinetics strongly influence the thermodynamic stability and phase morphology[239,242] (**Figure 4e**). In particular, reaction kinetics become rate limiting when diffusion is fast, and consequently the reaction rate rather than bulk thermodynamics determines whether phase separation occurs. This phenomenon was first predicted theoretically for $Li_xFePO_4$[207,239], where depth-averaged phase-field models revealed that phase separation occurring at equilibrium would be suppressed at high insertion rates. Rather than a static material property of a two-phase system, the solubility limits and spinodal points are dynamic properties controlled by the applied current. Additionally, the shape of the reaction rate curve (i.e. exchange current) with respect to composition can either enhance or suppress phase separation[195,233]. The suppression of phase separation at high rates has been confirmed experimentally for $Li_xFePO_4$[233,249–252], and electrochemical control of phase separation processes is now recognized as a general phenomenon that applies to pattern formation in many driven systems beyond battery electrodes[194,195,253].

Finally, at the electrode or device scale, phase-separating single crystals and polycrystalline agglomerates interact with each other. One manifestation of this is electrode-scale heterogeneity at equilibrium, also known as mosaic patterning, illustrated in **Figure 5d**. This phenomenon was rationalized by multiphase thermodynamics[153,254,255], which unified population dynamics models of single crystals[256] with porous electrode theory through non-equilibrium thermodynamics[194]. This process has been studied experimentally for lithium insertion into $Li_xFePO_4$[257,258] and $Li_xC_6$[223]. At equilibrium, an electrode of phase-separating crystals will consist of only homogenous crystals resting at energy minima (i.e., compositional bifurcation between particles). Any crystals at intermediate extent of insertion, and thus away from free energy minima, will experience a driving force to de/re-insert. Additionally, any crystals in a phase-separated state will be driven toward one of the single-phase energy minima by coherency strain energy[239] or surface energy[242]. Mosaic patterning runs counter to the behavior of many materials that homogenize at equilibrium. Small differences between crystals, such as their size, surface properties[242], or location in the electrode also play a role in determining the rate at which mosaic phase separation occurs as well as which crystals will undergo (de)insertion, and can even lead to open-circuit hysteresis[256]. Reaching a fully equilibrated mosaic state is possible in a porous electrode even when no current is applied, by intra- and inter-particle mechanisms such as the exchange of lithium through the electrolyte or surface diffusion[177].

## Outlook: ultrafast techniques

It is by now clear that ion insertion is enabled by, and in turn induces, material transformations across a broad range of length and time scales. Understanding across the breadth of involved scales and conceptual models (**Figures 3-5**) has been supported by a variety of established computational techniques from first-principles to continuum[16,259,260]. We now turn our attention to the shortest time scales associated with ion insertion and transport. The vibrational attempt frequencies ($\omega_0$ in **Figure 6**) for the hopping of many intercalating ions (e.g. $H^+$, $Li^+$, $Na^+$, $K^+$, $Ag^+$, and others) occur generally in the several-terahertz (THz) range, with periods shorter than a picosecond. In turn, cooperative responses to individual ionic hops involve dynamical heterogeneity and disorder evolving on picosecond timescales. Yet, nuclear magnetic resonance studies providing complementary elemental and chemical-environment resolution[261–263], and various modes of impedance spectroscopy[264], only reach nanosecond timescales. Recently-developed high-resolution electron microscopy[265–267] and synchrotron based x-ray (spectro)microscopy techniques enable coupled spatial and chemical characterization[18,268–270], but only at timescales of macroscopic devices.

Thus, the relationship between practical macroscopic observables, such as the Arrhenius activation energy for conductivity, and the microscopic material parameters, such as phononic structure, has remained a subject of active study since the 1950's[271–274]. The macroscopic ionic and polaronic conductivities are nontrivially linked to the vibrational modes involving the mobile ion, the local distortions accompanying hopping, and the picosecond-timescale dynamical heterogeneity[275]. The correlated motions of mobile ions[276,277] associated with fast ion conductivity[278,279], and collective ionic dynamics responsible for phase transitions[280], present particular interest.

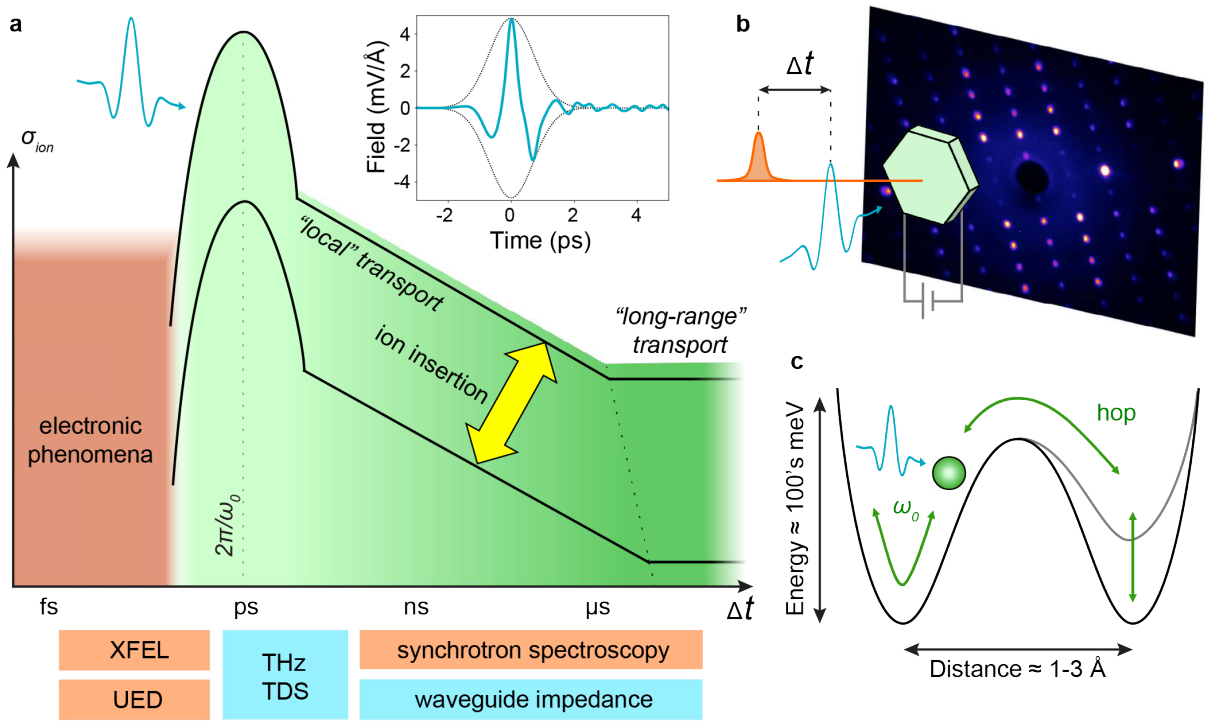

**Figure 6: Ultrafast time-domain probes of ion transport**. (**a**) Schematic of the vibrational and local ionic transport phenomena accessible to pump-probe experiments. Bottom: structural (orange) and optical (cyan) probe techniques at x-ray free-electron laser (XFEL) facilities, ultrafast electron diffraction (UED), terahertz time-domain spectroscopy (THz TDS), synchrotron, and impedance methods. Inset: temporal profile of an example near-single-cycle terahertz optical pulse obtained via optical rectification in lithium niobate. (**b**) Schematic of a pump-probe experiment: optical pump (cyan) and structural probe (orange). Transmission electron diffraction pattern provides a snapshot of dynamic structure. (**c**) Schematic of coupling an optical pump (cyan) to local ion transport (green): attempt frequency $\omega_0$, translation via a hop, and reorganization. Potential energy surfaces shown are metastable (grey) and relaxed (black). Diffraction pattern image in panel (b) adapted with permission from Ref. [285].

Looking ahead, we foresee unexplored opportunities to pump, control, and probe transport in ion insertion materials, including measurements of the ionic transition states and the associated non-equilibrium relaxations of the surrounding lattice cages[281,282]. Ultrafast techniques could enable the study and manipulation of ion motion at the atomic level and at timescales down to those associated with individual ionic and localized electronic hops (**Figure 2, Figure 6**). In the final section of this Review, we highlight the ongoing development of ultrafast characterization techniques. To probe such structural ionic dynamics at the timescales of individual transport events, at least two important challenges must be overcome.

First, one must find and make use of atomic-scale sensitive and high temporal resolution time-domain techniques, which access the nuclear rearrangements associated with ionic motion. In part, recently developed femtosecond x-ray and electron scattering approaches (**Figure 6b**) provide such an atomic-scale view of the underlying lattice motions[283–289]. However, much of the prior work in this field has focused on coherent responses to electronic excitations, where every unit cell responds roughly identically[280,284]. Meanwhile, ion hopping, transport, and insertion all occur in highly localized steps, where a minority of ions experiences the majority of the dynamics of interest within any given window of time. Furthermore, the dynamics themselves are incoherent and probabilistic because of the finite energy barrier separating ionic positions: even with a strong pump field, a diversity of responses can be anticipated. Additionally, in the fast-transport materials of practical interest with lower activation energies (e.g. < 200 meV), multiple pathways could contribute to overall transport, and dynamic disorder is enhanced. Thus, the development of time-domain diffuse scattering and pair distribution function approaches, which are sensitive to local nanoscale distortions acting as dynamic defects, is crucial towards progress here. For instance, measurements of the scattered x-ray intensity in the tails of Bragg peaks provide information on the elastic displacement field in the nanoscale region in the neighborhood of point defects [290–292]. Such studies could be usefully extended into the time domain using both femtosecond x-ray and electron diffuse scattering approaches, and to the practically relevant materials discussed in the preceding sections. Methods such as x-ray photon correlation spectroscopy[293,294] could further enable probing hopping dynamics and the associated structural fluctuations under both equilibrium and non-equilibrium conditions. Additional tabletop or operando approaches can also be usefully applied here.

The second key challenge is directly exciting ionic motion and hops as it pertains to the ion transport phenomena discussed above. Most pump-probe studies require a means to trigger a particular atomic distortion repeatedly and in the same way every time, in order to build up statistics on the measurement. Both interband[295] and intraband[296] electron transitions are readily pumped with tabletop ultrafast visible/near-IR frequency sources and can serve as such an impulsive trigger for vibrational dynamics[295]. In turn, terahertz time-domain spectroscopy[297,298] provides a way to probe materials on a picosecond timescale (**Figure 6a**). When coupled with e.g. a visible pump providing electronic excitation, terahertz spectroscopy presents a contact-free approach to measuring the mobilities of charge carriers[299–301], and has recently been applied to an ionic conductor in equilibrium[302]. However, one would more preferably find ways to probe ionic response under the influence of quasi-DC electric fields or with more narrowband pump pulses resonant with the hopping attempt frequencies. To generate such perturbations, one may make use of high-field, single-cycle pulses at THz frequencies[303–305] (inset in **Figure 6a**) that act on charge carriers as classical electromagnetic fields, and may further be quasi-resonant with ionic vibrational frequencies[306–308]. Such single- or few-cycle THz pulses can be thought of as all-optical or electrode-less biases to drive ions and electrons in specific directions defined by the polarization of the light field. After an applied field triggers a coherent ionic displacement (**Figure 6c**), the subsequent structural deformations can be directly visualized via changes in Bragg and diffuse scattering, or via coherent diffractive imaging. A number of examples exist in the literature demonstrating the use of THz fields to directly drive ions in materials[285,307–311]. For an order-of-magnitude estimate, an electric field of 10 MV/cm would move a $Na^+$ ion by ≈5 Å if applied over 0.5 ps in free space. One may view these measurements as a kind of ultrafast ion impedance spectroscopy, where one is probing not the long-range ionic and electronic conductivity, but rather the coupling between the vibrational modes of the lattice and the local ionic conductivity[264] (**Figure 6a**). Thus, high-field optical pulses in concert with theory and simulations could enable the impulsive excitation of coherent ionic dynamics and studies of both the atomistic nature of activation for ion transport, and of its modulation by ion insertion. Field strengths of ~10 MV/cm would also generate perturbations to interatomic potentials on the scale of tens of millivolts per angstrom, comparable with the activation energies of transport for ions and localized electrons. However, such field strengths are at the limit of what most tabletop sources can produce[304], and additional field-driven effects, most obviously heating, may arise in this strong field regime.

Following an ultrafast impulsive perturbation with a well-defined time zero point, a bevy of structural, spectroscopic, or electronic methods, up to and including lock-in current detection[312], can characterize the response of materials at all timescales. Further multimodal studies will enable selectivity and precision. Notably, bulk materials, interfaces, and junctions could be pumped and probed selectively. Several techniques could also impart spatial resolution on the kinetic timescales of ion insertion: transient gratings[313], introduction of an interface[89], introduction of a point scatterer such as a scanning probe[314], or thinning samples to nanoscale dimensions, such as for high-energy electron diffraction. Spatial resolution becomes available when employing x-ray dichroism resulting from magnetization as a contrast mode in x-ray microscopy[315–317]. The combination of electrochemistry with pump-probe methods to study the transport of species beyond electrons remains an attractive opportunity. New models drawing on phonon and solid-state physics, defect chemistry, and multi-body frameworks will be required to comprehensively describe these phenomena. Understanding the coupling between extended and localized vibrational modes and carrier conduction in practical systems is critical towards enabling the next generation of efficient devices for energy storage, catalysis, and computing.

## Conclusion

Although ion insertion has been studied extensively for energy storage and indeed powers a nearly $50 billion/year industry, opportunities to use ion insertion for both fundamental science and practical applications abound. The ability to precisely and dynamically tune material properties, often at room

temperature, extends the use of ion insertion to applications ranging from smart windows to neuromorphic computing to water desalination. While we have chosen to focus on the tuning of optical, electronic and thermal properties, these general principles have also been employed to modulate magnetic[318] and mechanical properties (i.e. actuation)[319,320]. Given their versatility and commercial appeal, we expect a surge in interest towards functional ion-insertion materials in the coming years. It is likely too, that new applications of the material-tuning aspect, which have not yet been considered, will be found. These could include examples of multifunctionality, i.e. multiple properties being tuned at the same time via ion insertion. For instance, efficient thermoelectrics require a combination of reduced thermal conductivity and enhanced electronic conductivity – a combination that can potentially be achieved by ion insertion. The fundamental science behind ion insertion remains complex due to the many simultaneously occurring processes which occur over a wide range of time and length scales. Recent years have seen a suite of novel experimental techniques developed to understand insertion processes with breathtaking spatial and chemical detail. Moving forward, we anticipate strong interest in driving and probing ion transport at ultrafast timescales approaching the fundamental modes of atomic vibrations, as well as at the ultralong timescales spanning the device lifetime. Ultimately, scientific understanding established in different application areas must be unified to facilitate the use of ion insertion in both established and emerging fields.

**Acknowledgements**

This work was supported by the U.S. Department of Energy, Office of Basic Energy Sciences, Division of Materials Sciences and Engineering (contract DE-AC02-76SF00515), as well as by the Toyota Research Institute through D3BATT: Center for Data-Driven Design of Li-Ion Batteries. PMC acknowledges support through the Stanford Graduate Fellowship as a Winston and Fu-Mei Chen fellow and through the National Science Foundation Graduate Research Fellowship under Grant No. DGE-1656518. We thank Y. Li (Univ. of Michigan), A. Salleo, Y. Cui, H. Thaman, A. Baclig, and A. Liang (Stanford Univ.), and M. Aykol (Toyota Research Institute) for discussions.

**Author Contributions**

All authors contributed to the discussion of content and editing of the manuscript prior to submission. AS led the organization and drafting of the manuscript, and wrote the sections on dynamic switching. ADP wrote the section on ultrafast techniques with inputs from AML. DAC wrote the sections on interfacial effects and phase transformations with inputs from DF. PMC wrote the section on point defects. JTM wrote the section on electrocatalysis. AS, ADP, DAC, PMC and JTM contributed equally and are listed alphabetically. DF wrote the section on CIET. MZB wrote the section on desalination. MFT, AML, MZB and WCC supervised the project and provided critical feedback on the draft. WCC conceptualized the Review.

**Competing Interests**

The authors declare no competing interests.